\documentclass{elsart}

\usepackage{graphicx}

\usepackage{amssymb}

\def\Dm#1#2{\Delta m_{#1#2}^2}
\def\Datm{\Delta m_{atm}^2}
\def\Dsol{\Delta m_{sol}^2}
\def\Dcho{\Delta m_{chooz}^2}

\def\sinatm{\sin^22\theta_{atm}}
\def\sinsol{\sin^22\theta_{sol}}
\def\sincho{\sin^22\theta_{chooz}}
\def\e3s{\eta^2_{e3}}
\def\ka#1#2{\frac{2\kappa_#1\kappa_#2}{\kappa_#1^2-\kappa_#2^2}}
\def\kb#1#2{\frac{\kappa_#1^2+\kappa_#2^2}{\kappa_#1^2-\kappa_#2^2}}
\def\tep{\tilde\epsilon}
\def\PRL#1#2#3{{\em Phys. Rev. Lett.\/} {\bf #1} (#2) #3}
\def\PRC#1#2#3{{\em Phys. Rev.\/} {\bf C#1} (#2) #3}
\def\PRD#1#2#3{{\em Phys. Rev.\/} {\bf D#1} (#2) #3}
\def\PLB#1#2#3{{\em Phys. Lett.\/} {\bf B#1} (#2) #3}
\def\NPB#1#2#3{{\em Nucl. Phys.\/} {\bf B#1} (#2) #3}

\def\EPJC#1#2#3{{\em Eur. Phys. J.\/} {\bf C#1} (#2) #3}
\def\AJ#1#2#3{{\em Astrophys. J.\/} {\bf #1} (#2) #3}

\begin{document}

\begin{frontmatter}




\title{Model for neutrino mixing based on SO(10)}


\author{Noriyuki Oshimo}

\address{
Institute of Humanities and Sciences {\rm and} Department of Physics \\
Ochanomizu University, Tokyo, 112-8610, Japan
} 

\begin{abstract}

     Assuming grand unified theory (GUT) and supersymmetry,
we propose a simple model which can consistently accommodate the
masses and mixings for quarks and leptons.
The grand unified group is SO(10), and $\bf 10$, $\bf 120$, 
and $\overline {\bf 126}$ representations are introduced for 
the Higgs superfields which give masses to the quarks and leptons.
The differences of masses and mixings between the quarks and the leptons
are attributed to the Higgs boson structure.
Below the GUT energy scale, the model is the same as the minimal
supersymmetric standard model except its inclusion of dimension-5
operators for the small neutrino masses.
The renormalization group equations of the independent parameters
for the Higgs couplings with the quarks and leptons  
are given explicitly to connect the two energy scales of GUT and
electroweak theory.

\end{abstract}

\begin{keyword}
neutrinos \sep SO(10) \sep Higgs 


\PACS 12.10.Dm \sep 12.15.Ff \sep 12.60.Jv \sep 14.60.Pq 

\end{keyword}
\end{frontmatter}


\section{Introduction}
\label{}

     Implications of non-vanishing neutrino masses are accumulating from 
the experiments for solar \cite{sol} and atmospheric \cite{atm} neutrinos 
which show neutrino oscillations.   
This is the first experimental evidence that suggests physics beyond  
the standard model (SM).   
The possibility of oscillation has also been examined 
by the neutrinos from nuclear reactors.  The negative result of 
CHOOZ~\cite{chooz} gives 
constraints on the masses and mixing allowed for the neutrinos.  
The result of KamLAND \cite{kamland} confirms the oscillations of 
the solar neutrinos.  
It is a task for the extension of the SM to accommodate these 
experimental results.  
In particular, the extreme smallness of the masses and 
the largeness of the generation mixing angles, 
both of which are observed in the experiments, 
should be accounted for naturally.  

     From a theoretical point of view, it is reasonable for the SM 
to be extended under grand unified theory (GUT).   
Furthermore, supersymmetry may have to be introduced 
in order that the SM be consistently embedded in GUT models.  
On the other hand, some of the GUT models, such as those based on 
SO(10)~\cite{so10}, automatically contain right-handed neutrinos.  
If they have large Majorana masses, 
the lightness of the neutrinos could be naturally understood.  
The plausibility of supersymmetric GUT models as the extension of the SM 
are then further strengthened.     

     One tough obstacle confronts GUT models, concerning 
the generation mixing of the leptons.  
The lepton mixing is described by the Maki-Nakagawa-Sakata (MNS) matrix 
as the Cabibbo-Kobayashi-Maskawa (CKM) matrix does the quark mixing.   
Contrary to the CKM matrix, the observed neutrino oscillations show 
that the off-diagonal elements of the MNS matrix are not suppressed.  
However, since the leptons and the quarks are contained in the same 
representations of the grand unified group, similar magnitudes are 
deduced, by simple consideration, for corresponding elements of 
the two matrices.   
A successful GUT model is required to naturally implement the difference of 
mixing between the quarks and the leptons.  
Although there are various solutions advocated for the 
problem \cite{dorsner}, rather contrived schemes have been invoked.  

     Aiming at a natural explanation for 
the generation mixings of the leptons and the quarks, 
we present a model base on SO(10) and supersymmetry in which the difference 
is simply attributed to the Higgs boson structure \cite{oshimo}.  
This model includes $\bf 10$, $\bf 120$, and $\overline {\bf 126}$ 
representations of the SO(10) group for the Higgs bosons 
which give masses to the quarks and leptons.  
The $\bf 120$ representation becomes the origin of the different 
generation mixings.  
Large Majorana masses for the right-handed neutrinos are generated 
by a vacuum expectation value (VEV) of the $\overline {\bf 126}$ 
representation.  
Below the GUT energy scale, this model is described as the minimal 
supersymmetric standard model (MSSM) with dimension-5 operators 
composed of the SU(2) doublet superfields for the left-handed leptons 
and the Higgs bosons.  
The coefficients of the Higgs couplings with the quarks and leptons 
evolve between the energy scales of GUT and electroweak theory.  
We obtain the renormalization group equations for the eigenvalues 
of the coefficient matrices and the independent arguments of the 
MNS and CKM matrices.  
Quantitative analyses are then made on the neutrino oscillations.  
It is shown that the lepton mixing observed is compatible with 
the quark mixing in certain ranges of the model parameters.   

     In SO(10) GUT models the Higgs bosons for the quark and 
lepton masses must be contained in $\bf 10$, $\bf 120$, or  
$\overline {\bf 126}$ representations.  
The direct product for one of these three representations with 
two spinor $\bf 16$ representations for the quarks and leptons 
becomes an SO(10) singlet.  
The model with two Higgs fields of $\bf 10$ and $\overline {\bf 126}$ 
has been studied extensively \cite{babu}.  
However, all the experimental results for the neutrino 
oscillations are only marginally accommodated.   
The $\bf 10$ and $\overline{\bf 126}$ representations make the Higgs 
couplings symmetric on the generation indices, and their SU(2)-doublet 
components couple to both the quarks and the leptons. 
On the other hand, the $\bf 120$ representation leads to 
antisymmetric Higgs couplings, and the SU(2)-doublet components which  
couple to the quarks may be different from those to the leptons.  
By including $\bf 120$, the correlation of the masses and mixings  
between the quarks and the leptons becomes more flexible. 
 
     In sect. 2 we summarize the interpretation of neutrino oscillations 
in the framework of the model with three generations for leptons.  
In sect. 3 the renormalization group equations for the MSSM with 
the dimension-5 operators  
are obtained explicitly in terms of independent parameters for 
the coefficients of Higgs couplings with quarks and leptons.   
In sect. 4 the GUT based on SO(10) and supersymmetry is discussed.  
After deriving in general the relations of Higgs couplings for the quarks 
and the leptons, a plausible model is specified.  
In sect. 5 we make numerical analyses of the masses and 
mixing of the neutrinos.  
Conclusions are given in sect. 6.   
 
\section{Neutrino Oscillations}

     The deficit of neutrinos from the sun and the anomaly for  
neutrinos from the atmosphere can be explained by neutrino oscillations.     
The atmospheric neutrino observation gives that the mass-squared difference 
and the mixing angle between the $\mu$-neutrino and a certain  
neutrino are given by  
\begin{equation}
 \Datm = (1-8)\times10^{-3} {\rm eV^2},  \quad  
 \sinatm > 0.85.  
\label{expatm}
\end{equation}
There are several oscillation  scenarios for the solar neutrino problem.  
However, a most likely solution is given by a large mixing angle 
between the $e$-neutrino and a certain neutrino under 
the Mikheyev-Smirnov-Wolfenstein effect, suggesting the ranges 
\begin{equation}
 \Dsol = (2-10)\times10^{-5} {\rm eV^2},  \quad   
 \sinsol = 0.5-0.9.  
\label{expsol}
\end{equation}
The measurements by KamLAND are consistent with these results.  
On the other hand, by the experiment of CHOOZ the oscillation has 
not been observed for the $e$-neutrino, which excludes the combined region  
\begin{equation}
 \Dcho > 1\times 10^{-3} {\rm eV^2},  \quad 
 \sincho > 0.2.   
\label{expchooz}
\end{equation}
These experimental measurements are taken into consideration in 
constructing our model.  

     We interpret the above results in the model with 
three generations of massive neutrinos.  
In this framework the strength of the $W$-boson interaction for a 
neutrino and a charged lepton depends on the generations.  
The mass eigenstate $\nu_i$ for the neutrino of the $i$-th generation 
is related to the eigenstate $\tilde\nu_j$ of interaction with 
the charged lepton of the $j$-th generation by 
\begin{equation}
      \nu_i = (V_{MNS})_{ij}\tilde\nu_j,  
\end{equation}
where $V_{MNS}$ denotes the MNS matrix.  
The survival probability for the interaction eigenstate $\tilde\nu_i$ 
after run of distance $L$ with energy $E$ is given by 
\begin{equation}
 P(\tilde\nu_i \to \tilde\nu_i) = 
\biggl|\sum_{k=1}^3|(V_{MNS})_{ki}|^2
\exp(-i\frac{m_{\nu k}^2}{2E}L)\biggr|^2,   
\label{probability}   
\end{equation}
where $m_{\nu k}$ represents the mass eigenvalue for the neutrino 
of the $k$-th generation.  
The mass-squared difference is hereafter written as 
\begin{equation}
\Delta m_{ij}^2=m_{\nu i}^2-m_{\nu j}^2.  
\end{equation}
We assume that the neutrino masses are hierarchical, 
$m_{\nu 1} < m_{\nu 2} < m_{\nu 3}$, and not degenerated, similarly to the 
quarks and charged leptons.

     In the CHOOZ experiment, 
the phases in Eq. (\ref{probability}) are found to satisfy    
the relations $(\Dm21/2E)L \ll (\Dm31/2E)L \sim 1$.   
The survival probability of the $e$-neutrino is expressed by 
\begin{equation}
 P(\tilde\nu_e \to \tilde\nu_e) = 
1-4|(V_{MNS})_{31}|^2(1-|(V_{MNS})_{31}|^2)\sin^2\frac{\Dm31}{4E}L.  
\end{equation} 
Thus, the measured quantities are translated as  
\begin{equation}
    \Dcho   = \Dm31,  \quad    
    \sincho = 4|(V_{MNS})_{31}|^2(1-|(V_{MNS})_{31}|^2).  
\label{sincho}
\end{equation}
For the atmospheric neutrino oscillation, the phases in 
Eq. (\ref{probability}) satisfy the relations 
$(\Dm21/2E)L \ll (\Dm32/2E)L$ and $(\Dm21/2E)L \ll 1$.  
The survival probability of the $\mu$-neutrino is given by 
\begin{equation}
 P(\tilde\nu_\mu \to \tilde\nu_\mu) = 
1-4|(V_{MNS})_{32}|^2(1-|(V_{MNS})_{32}|^2)\sin^2\frac{\Dm32}{4E}L.  
\end{equation} 
The experimental results are expressed by 
\begin{equation}
    \Datm   = \Dm32,  \quad   
    \sinatm = 4|(V_{MNS})_{32}|^2(1-|(V_{MNS})_{32}|^2).  
\label{sinatm}
\end{equation}
For the solar neutrino oscillation, the relations 
$|(V_{MNS})_{31}|^2 \ll |(V_{MNS})_{21}|^2$ and 
$|(V_{MNS})_{31}|^2 \ll |(V_{MNS})_{11}|^2$ could hold 
from Eqs. (\ref{expatm}), (\ref{expchooz}), (\ref{sincho}), 
and (\ref{sinatm}).  
These constraints make it possible to evaluate the survival 
probability of the $e$-neutrino by  
\begin{equation}
 P(\tilde\nu_e \to \tilde\nu_e) = 
1-4|(V_{MNS})_{21}|^2(1-|(V_{MNS})_{21}|^2)\sin^2\frac{\Dm21}{4E}L.  
\end{equation} 
Therefore, we obtain the equations  
\begin{equation}
    \Dsol = \Dm21,  \quad   
    \sinsol = 4|(V_{MNS})_{21}|^2(1-|(V_{MNS})_{21}|^2).  
\end{equation}
These evaluations are used to discuss the predictions of the model.   

\section{Energy evolution}

     We assume that physics below the GUT energy scale is 
described by the MSSM with the dimension-5 operators composed of the  
superfields for left-handed leptons and Higgs bosons.  
This assumption is fulfilled in the model which is presented afterward.  
The superpotential relevant to the quark and lepton 
masses is given by  
\begin{eqnarray}
 W  &=& W_1 + W_2,  \nonumber \\
W_1 &=& \eta_d^{ij} H_1\times Q^iD^{cj} + \eta_u^{ij}H_2\times Q^iU^{cj}
 + \eta_e^{ij} H_1\times L^iE^{cj} + {\rm H.c.},  \label{superpotential1} \\
W_2 &=& \frac{1}{2}\kappa^{ij}H_2\times L^iH_2\times L^j + {\rm H.c.}, 
\label{superpotential2}
\end{eqnarray}
where $H_1$ and $H_2$ stand for the superfields for the Higgs bosons 
with hypercharges $-1/2$ and 1/2, respectively.  
Superfields are denoted by $Q^i$, $U^{ci}$, and $D^{ci}$ for the quarks and 
$L^i$ and $E^{ci}$ for the leptons, in a self-explanatory notation, 
with $i$ being the generation index.  
The SU(3) group indices are understood.  
The dimension-5 superpotential $W_2$ yields the operators 
\begin{eqnarray}
   L &=& -\frac{1}{2}\kappa^{ij}
  \left(\begin{array}{c} \phi_{H_2}^+ \\ 
                         \phi_{H_2}^0 \end{array} \right) \times  
  \left(\begin{array}{c} \overline{(\nu_L^i)^c} \\ 
                         \overline{(e_L^i)^c} \end{array} \right)  
  \left(\begin{array}{c} \phi_{H_2}^+ \\ 
                         \phi_{H_2}^0 \end{array} \right) \times  
  \left(\begin{array}{c} \nu_L^{j} \\ 
                         e_L^{j} \end{array} \right)  
 + {\rm H.c.},  
\label{dimension5}
\end{eqnarray}
where $\phi_{H_2}^+$ and $\phi_{H_2}^0$ represent the scalar components 
of the superfield $H_2$.   
At the electroweak energy scale where the SU(2)-doublet Higgs bosons 
have non-vanishing VEVs, these dimension-5 operators become tiny 
Majorana mass terms for the left-handed neutrinos.  
The other operators arising from $W_2$ cause negligible effects.   

     The coefficient matrices $\eta_d$, $\eta_u$, $\eta_e$, and $\kappa$ 
are diagonalized by unitary matrices $U_L$'s and $U_R$'s.   
The CKM matrix for the quarks and the MNS matrix for the leptons are 
defined by 
\begin{eqnarray}
V_{CKM} &=& U_L^{u\dagger}U_L^d,  \quad    
   U_L^{uT}\eta_uU_R^{u*}=\eta^D_u, \quad 
 U_L^{dT}\eta_dU_R^{d*}=\eta^D_d,  \\
V_{MNS} &=& U_L^{\nu\dagger}U_L^e, \quad    
  U_L^{\nu T}\kappa U_L^{\nu}=\kappa^D, \quad 
 U_L^{eT}\eta_eU_R^{e*}=\eta^D_e,  
\end{eqnarray}
where $\eta_d^D$, $\eta_u^D$, $\eta_e^D$, and $\kappa^D$ 
stand for diagonalized matrices.  
Taking the VEVs of electroweak symmetry breaking for positive, 
the diagonal elements of $\eta_d^D$, $\eta_e^D$, and $\kappa^D$ 
should be positive, while those of $\eta_u^D$ should be negative.   
The quarks and leptons then have positive masses.  
Those diagonal elements are expressed by $\eta_{di}$, $\eta_{ui}$, 
$\eta_{ei}$, and $\kappa_i$.  

     A 3$\times$3 unitary matrix has nine independent parameters.  
For the explicit expression of $V_{CKM}$ or $V_{MNS}$, we adopt the 
following parametrization \cite{naculich}:  
\begin{eqnarray}
       V &=& P_+ V_0P_-,  
\label{unitary_matrix} \\
       V_0 &=& \left(
        \begin{array}{ccccc}
 s_1s_2c_3+{\rm e}^{i\delta}c_1c_2 & & c_1s_2c_3-{\rm e}^{i\delta}s_1c_2 & &  
                                 s_2s_3  \\
 s_1c_2c_3-{\rm e}^{i\delta}c_1s_2 & & c_1c_2c_3+{\rm e}^{i\delta}s_1s_2 & &  
                                 c_2s_3  \\
 -s_1s_3  & & -c_1s_3 & & c_3
        \end{array}
        \right),  \nonumber \\
  P_+ &=& {\rm diag}(\exp(i\rho_1),\exp(i\rho_2),1), 
                                           \nonumber \\
  P_- &=& {\rm diag}(\exp(-i\rho_3),\exp(-i\rho_4),\exp(-i\rho_5)), 
                                           \nonumber 
\end{eqnarray}
where $c_i=\cos\theta_i$ and $s_i=\sin\theta_i$ ($i=1,2,3$).  
Without loss of generality, the angles $\theta_1$, $\theta_2$, 
and $\theta_3$ can be taken to lie in the first quadrant.  
At the electroweak energy scale, the numbers of the physical parameters 
for the CKM matrix and the MNS matrix are four and six, respectively.  
However, we need the general form of parametrization 
in Eq. (\ref{unitary_matrix}) 
to discuss the energy dependencies of the CKM and MNS matrices.   

     The values of $\eta^D_d$, $\eta^D_u$, $\eta^D_e$, $\kappa^D$, 
$V_{CKM}$, and $V_{MNS}$ evolve depending on the energy scale.  
The renormalization group equations for these parameters  
and the gauge coupling constants of SU(3)$\times$SU(2)$\times$U(1) 
close on themselves at the one-loop level.  
We give the equations of the independent parameters, where 
the inequalities 
$\eta^2_{da}\ll \eta^2_{d3}$, $\eta^2_{ua}\ll \eta^2_{u3}$, and 
$\eta^2_{ea}\ll \eta^2_{e3}$ ($a=1,2$) are taken into account.  \\ 
{\it The gauge coupling constants:}
\begin{equation}
      \mu\frac{d g_3^2}{d\mu} = -3\frac{g_3^4}{8\pi^2},  
                           \quad 
      \mu\frac{d g_2^2}{d\mu} = \frac{g_2^4}{8\pi^2},  
                           \quad 
      \mu\frac{d g_1^2}{d\mu} = \frac{33}{5}\frac{g_1^4}{8\pi^2}.  
\end{equation}
{\it The diagonalized Higgs coupling coefficients:}
\begin{eqnarray}
\mu\frac{d\eta^2_{di}}{d\mu} &=& -\frac{\eta^2_{di}}{8\pi^2}
         \biggl[\frac{16}{3}g_3^2+3g_2^2+\frac{7}{15}g_1^2
     -3\eta^2_{d3}-3\eta^2_{di}-\eta^2_{e3}   \nonumber \\
       & &   -|\left(V_{CKM}\right)_{3i}|^2\eta^2_{u3}\biggr], 
                           \\
\mu\frac{d\eta^2_{ui}}{d\mu} &=& -\frac{\eta^2_{ui}}{8\pi^2}
         \biggl[\frac{16}{3}g_3^2+3g_2^2+\frac{13}{15}g_1^2
            -3\eta^2_{u3}-3\eta^2_{ui}    \nonumber \\
       & &  -|\left(V_{CKM}\right)_{i3}|^2\eta^2_{d3}\biggr], 
                            \\
\mu\frac{d\eta^2_{ei}}{d\mu} &=& -\frac{\eta^2_{ei}}{8\pi^2}
         \left[3g_2^2+\frac{9}{5}g_1^2
                  -\eta^2_{e3}-3\eta^2_{ei}-3\eta^2_{d3}\right], 
                            \\
\mu\frac{d\kappa_i}{d\mu} &=& -\frac{\kappa_i}{8\pi^2}
         \left[3g_2^2+\frac{3}{5}g_1^2
                  -3\eta^2_{u3}  
         -|\left(V_{MNS}\right)_{i3}|^2\eta^2_{e3}\right]. 
\end{eqnarray}
{\it The CKM matrix:}
\begin{eqnarray}
\mu\frac{d\theta_{q1}}{d\mu} &=& -\frac{\eta^2_{u3}}{16\pi^2}s_1c_1s_3^2, 
                       \\
\mu\frac{d\theta_{q2}}{d\mu} &=& -\frac{\eta^2_{d3}}{16\pi^2}s_2c_2s_3^2, 
                        \\
\mu\frac{d\theta_{q3}}{d\mu} &=& -\frac{\eta^2_{u3}
                                 +\eta^2_{d3}}{16\pi^2}s_3c_3, 
                        \\
\mu\frac{d\delta_q}{d\mu} &=& 0,   
                      \\
\mu\frac{d\rho_{qa}}{d\mu} &=& 0    \quad \quad  (a=1-5)  ,   
\end{eqnarray}
where $\theta_{qi}$ and $\delta_q$ stand for the arguments in $V_0$ 
for the CKM matrix, with $c_i=\cos\theta_{qi}$ and $s_i=\sin\theta_{qi}$, 
and $\rho_{qa}$ represents an additional phase parameter 
in $P_+$ and $P_-$.   
                            \\
{\it The MNS matrix:}
\begin{eqnarray}
\mu\frac{d\theta_{l1}}{d\mu} &=& -\frac{\eta^2_{e3}}{16\pi^2}s_2c_2c_3
        \left[(B_R+C_R)c_\delta-(B_I-C_I)s_\delta\right], 
                        \\
\mu\frac{d\theta_{l2}}{d\mu} &=& -\frac{\eta^2_{e3}}{16\pi^2}s_2c_2
         \left[-A_Rs_3^2+(B_R+C_R)c_3^2\right], 
                        \\
\mu\frac{d\theta_{l3}}{d\mu} &=& -\frac{\eta^2_{e3}}{16\pi^2}s_3c_3
         \left[-B_Rc_2^2+C_Rs_2^2\right], 
                        \\
\mu\frac{d\delta_l}{d\mu} &=& -\frac{\eta^2_{e3}}{16\pi^2}
 \left[-A_Is_3^2-(B_R+C_R)s_\delta\frac{c_1^2-s_1^2}{s_1c_1}s_2c_2c_3 \right.  
                          \nonumber \\ 
   & & \left.  - B_I\left(c_\delta\frac{c_1^2-s_1^2}{s_1c_1}s_2c_2c_3
                       +s_2^2c_3^2-c_2^2\right) \right.  \nonumber \\
 & &  \left. + C_I\left(c_\delta\frac{c_1^2-s_1^2}{s_1c_1}s_2c_2c_3
                       -c_2^2c_3^2+s_2^2\right)\right], 
                        \\
\mu\frac{d\rho_{l1}}{d\mu} &=& -\frac{\eta^2_{e3}}{16\pi^2}
                 \left[(A_I-B_I)c_2^2s_3^2+C_I(c_3^2-s_2^2s_3^2)\right], 
                        \\
\mu\frac{d\rho_{l2}}{d\mu} &=& -\frac{\eta^2_{e3}}{16\pi^2}
                     \left[(A_I-C_I)s_2^2s_3^2+B_I(c_3^2-c_2^2s_3^2)\right], 
                        \\
\mu\frac{d\rho_{l3}}{d\mu} &=& \frac{\eta^2_{e3}}{16\pi^2}c_3
 \left[(B_R+C_R)s_\delta\frac{c_1}{s_1}s_2c_2 \right.  
                          \nonumber \\ 
   & & + \left. B_I\left(c_\delta\frac{c_1}{s_1}s_2c_2-c_2^2c_3\right) 
-C_I\left(c_\delta\frac{c_1}{s_1}s_2c_2+s_2^2c_3\right)\right], 
                        \\
\mu\frac{d\rho_{l4}}{d\mu} &=& \frac{\eta^2_{e3}}{16\pi^2}c_3
 \left[-(B_R+C_R)s_\delta\frac{s_1}{c_1}s_2c_2 \right.  
                          \nonumber \\ 
   & & - \left. B_I\left(c_\delta\frac{s_1}{c_1}s_2c_2+c_2^2c_3\right) 
+C_I\left(c_\delta\frac{s_1}{c_1}s_2c_2-s_2^2c_3\right)\right], 
                        \\
\mu\frac{d\rho_{l5}}{d\mu} &=& \frac{\eta^2_{e3}}{16\pi^2}s_3^2
                 \left[B_Ic_2^2+C_Is_2^2\right], 
                        \\
A_R &=& \ka12\cos 2(\rho_1-\rho_2)+\kb12,  
                   \nonumber \\
A_I &=& \ka12\sin 2(\rho_1-\rho_2),  
                  \nonumber \\
B_R &=& \ka23\cos 2\rho_2+\kb23,  
                   \nonumber \\
B_I &=& \ka23\sin 2\rho_2,  
                \nonumber \\
C_R &=& \ka31\cos 2\rho_1+\kb31,  
                       \nonumber \\
C_I &=& -\ka31\sin 2\rho_1,  
                    \nonumber  
\end{eqnarray}
where $\theta_{li}$ and $\delta_l$ stand for the arguments in $V_0$ 
for the MNS matrix, with $c_i=\cos\theta_{li}$, $s_i=\sin\theta_{li}$, 
$c_\delta=\cos\delta_l$, and $s_\delta=\sin\delta_l$, and 
$\rho_{la}$ represents an additional phase parameter in $P_+$ and $P_-$.   
The mutual dependencies of the independent parameters in energy 
evolution are manifestly 
seen, thanks to the explicit expressions in terms of the parameters 
themselves.     
For instance, the MNS matrix receives large quantum corrections 
if some of the neutrino mass coefficients $\kappa_i$ are roughly 
degenerated \cite{haba}.   

\section{Model}

     The grand unified group of our model is SO(10).  
Its spinor $\bf 16$ representation contains all the 
superfields for quarks and leptons of one generation, 
for both left-handed and right-handed components.  
The right-handed neutrinos are therefore naturally incorporated.  
The decomposition of the direct product for two $\bf 16$'s is given by  
${\bf 16}\times{\bf 16}={\bf 10}+{\bf 120}+{\bf 126}$.  
The Higgs superfields which give masses to the quarks and 
leptons must be assigned to $\bf 10$, $\bf 120$, or $\overline{\bf 126}$ 
representations.  
We introduce one superfield for each representation.  

     The superpotential relevant to the quark and lepton masses 
are given, in the framework of SU(3)$\times$SU(2)$\times$U(1), by  
\begin{eqnarray}
W &=& \eta^{ij}\left[H^{\bar 5}_{10}\times\left(Q^iD^{cj}+L^iE^{cj}\right)
             +H^5_{10}\times\left(Q^iU^{cj}+L^iN^{cj}\right)\right] 
                              \nonumber \\
&+& \epsilon^{ij}\left[H^{\bar 5}_{120}\times\left(Q^iD^{cj}+L^iE^{cj}\right)
              +\frac{1}{\sqrt{3}}H^{\overline{45}}_{120}\times
                        \left(Q^iD^{cj}-3L^iE^{cj}\right) \right. 
                                   \nonumber \\
 & & \left.   +2H^5_{120}\times L^iN^{cj} 
            +\frac{2}{\sqrt{3}}H^{45}_{120}\times Q^iU^{cj}\right] 
                              \nonumber \\
&+& \zeta^{ij}\left[H^{\overline{45}}_{\overline{126}}\times
       \left(Q^iD^{cj}-3L^iE^{cj}\right) 
            +H^5_{\overline{126}}\times\left(Q^iU^{cj}-3L^iN^{cj}\right) 
                             \right.    \nonumber \\
 & & \left. +\sqrt{6}L^{iT}H^{15}_{\overline{126}}L^j 
            +\sqrt{6}H^1_{\overline{126}}N^{ci}N^{cj} \right] + {\rm H.c.},
\label{couplings}
\end{eqnarray}
where superfields $N^{ci}$ for the right-handed neutrinos 
appear in addition to the superfields $Q^i$, $U^{ci}$, $D^{ci}$, 
$L^i$, and $E^{ci}$ for the quarks and leptons.   
Higgs superfields are denoted by $H$'s with upper and lower indices 
showing transformation properties under SU(5) and SO(10), respectively:  
$H^{\bar 5}_{10}$, $H^{\bar 5}_{120}$, $H^{\overline{45}}_{120}$, 
and $H^{\overline{45}}_{\overline{126}}$ are SU(2) doublets with 
hypercharge $-1/2$; 
$H^5_{10}$, $H^5_{120}$, $H^{45}_{120}$, and 
$H^5_{\overline{126}}$ are SU(2) doublets with hypercharge 1/2; 
$H^{15}_{\overline{126}}$ is an SU(2) triplet; 
and $H^1_{\overline{126}}$ is an SU(2)$\times$U(1) singlet.   
Each superfield has been normalized.   
Owing to the antisymmetric property of $\bf 120$, the coupling 
$H^5_{120}\times Q^iU^{cj}$ does not appear.  
The coupling constants $\eta^{ij}$ and $\zeta^{ij}$ are symmetric for the
generation indices, while $\epsilon^{ij}$ are antisymmetric.

     We can see from Eq. (\ref{couplings}) the characteristic of 
the $\bf 120$ representation.  
Any SU(2) doublet in $\bf 10$ or $\overline{\bf 126}$ couples 
both the quark and the lepton superfields.  
As a result, in Eqs. (\ref{superpotential1}) and (\ref{superpotential2}), 
the Higgs coupling coefficients for the leptons become related 
to those for the quarks.   
On the other hand, for $\bf 120$, four SU(2) doublets 
$(\sqrt{3}H^{\bar 5}_{120}+H^{\overline{45}}_{120})/2$, 
$(H^{\bar 5}_{120}-\sqrt{3}H^{\overline{45}}_{120})/2$, 
$H^5_{120}$, and $H^{45}_{120}$ couple to 
$Q^iD^{cj}$, $L^iE^{cj}$, $L^iN^{cj}$, and $Q^iU^{cj}$, respectively.  
The Higgs coupling coefficients could be less constrained.  

     The SU(2)-doublet Higgs superfields for electroweak symmetry  
breaking are composed of the superfields in $\bf 10$, 
$\bf 120$, $\overline{\bf 126}$, and some other representations.  
Among the possible linear combinations of SU(2) doublets with the same 
hypercharge, only one doublet should be kept light to satisfy   
the unification of the gauge coupling constants for 
SU(3)$\times$SU(2)$\times$U(1).   
The two doublets with hypercharges $-1/2$ and $1/2$ assume the role 
of the Higgs superfields $H_1$ and $H_2$ in 
Eqs. (\ref{superpotential1}) and (\ref{superpotential2}).   
The other linear combinations must have large masses and decouple 
from theory below the GUT energy scale.
We express the Higgs superfields by
\begin{eqnarray}
H_1 &=& (C_1^\dagger)_{11}H^{\bar 5}_{10}
          +(C_1^\dagger)_{12}H^{\bar 5}_{120}
          +(C_1^\dagger)_{13}H^{\overline{45}}_{120}
          +(C_1^\dagger)_{14}H^{\overline{45}}_{\overline{126}}+...,   
\label{higgs_h1}   \\
H_2 &=& (C_2^\dagger)_{11}H^5_{10}
          +(C_2^\dagger)_{12}H^5_{120}
          +(C_2^\dagger)_{13}H^{45}_{120}
          +(C_2^\dagger)_{14}H^5_{\overline{126}}+...,
\label{higgs_h2}
\end{eqnarray}
where $C_1$ and $C_2$ represent unitary matrices.   
Some components of $H_1$ and $H_2$ belong to the representations 
different from $\bf 10$, $\bf 120$, and $\overline{\bf 126}$, 
which are denoted by the ellipses.  
For instance, one superfield of $\bf 126$ is included in the model.   
Its SU(5)-singlet scalar component has a large VEV to cancel 
the VEV of $H^1_{\overline{126}}$, and the VEVs of the auxiliary D 
fields for SO(10) are kept small.  
This $\bf 126$ representation contains SU(2) doublets, which may become 
the components of $H_1$ and $H_2$.       
 
    The matrices $C_1$ and $C_2$ should be determined by the Higgs potential 
at the GUT energy scale.  
However, we take them as independent parameters in this paper, 
assuming that an appropriate Higgs potential could be constructed.   
Since the Higgs potential contains also the fields 
which are to break SO(10) correctly down to SU(3)$\times$SU(2)$\times$U(1), 
it is very complicated to analyze the whole potential.  
The Higgs potential is also supposed to induce the well-known split 
between the light SU(2) doublets and the heavy SU(3) triplets, 
as well as the split between the light SU(2) doublets, $H_1$ and $H_2$,  
and the other heavy SU(2) doublets.   
 
     The superpotential in Eq. (\ref{superpotential1}) is now determined.  
The coefficient matrices are given by 
\begin{eqnarray}
\eta_d &=& \eta(C_1)_{11}+\epsilon[(C_1)_{21}+
              \frac{1}{\sqrt{3}}(C_1)_{31}]+\zeta(C_1)_{41}, 
                \label{etad} \\
\eta_u &=& \eta(C_2)_{11}+\frac{2}{\sqrt{3}}\epsilon(C_2)_{31}
                        +\zeta(C_2)_{41}, 
                \label{etau} \\
\eta_e &=& \eta(C_1)_{11}+\epsilon[(C_1)_{21}-\sqrt{3}(C_1)_{31}]
                            -3\zeta(C_1)_{41}.     
                \label{etae} 
\end{eqnarray}
The dimension-5 superpotential in Eq. (\ref{superpotential2}) is 
induced by the interactions of $N^{ci}$ in Eq. (\ref{couplings}),  
\begin{eqnarray}
W &=& \eta_\nu^{ij} H_2\times L^iN^{cj} 
     + \sqrt{6}\zeta^{ij}H^1_{\overline{126}}N^{ci}N^{cj} + {\rm H.c.},  
                         \\
\eta_\nu &=& \eta(C_2)_{11}+2\epsilon(C_2)_{21}-3\zeta(C_2)_{41}.  
     \label{etanu}    
\end{eqnarray}
If the scalar component of $H^1_{\overline{126}}$ has a large VEV,  
the right-handed neutrinos and sneutrinos receive large masses.   
The mass matrix of the neutrinos is given by 
\begin{equation}
  M_{\nu_R} = 2\sqrt{3}v_S\zeta,  
\end{equation}
with $<H^1_{\overline{126}}>=v_S/\sqrt{2}$.  
The mass-squared matrix of the sneutrinos is expressed as 
$M_{\nu_R}M_{\nu_R}^\dagger$.  
Exchanges of these particles lead to an effective superpotential 
given in Eq.~(\ref{superpotential2}), 
\begin{equation}
\kappa = -\eta_\nu\left(M_{\nu_R}\right)^{-1}\eta_\nu^T.  
\label{kappamatrix}   
\end{equation}
The coupling $L^{iT}H^{15}_{\overline{126}}L^j$ in Eq. (\ref{couplings})
could give Majorana masses to the left-handed neutrinos, 
if the neutral scalar component of $H^{15}_{\overline{126}}$ has a 
non-vanishing VEV.  
However, this VEV has to be as small as the neutrino masses measured 
in experiments.  
Then, an extreme fine-tuning of the Higgs potential would become inevitable.    
We therefore discard this possibility, assuming that 
$H^{15}_{\overline{126}}$ is heavy enough not to develop 
a non-vanishing VEV.  

     The coefficient matrices $\eta_d$, $\eta_u$, and $\eta_e$ in 
Eq. (\ref{superpotential1}) and $\kappa$ in Eq. (\ref{superpotential2}) 
at the GUT energy scale are related to each other through 
Eqs. (\ref{etad}), (\ref{etau}), (\ref{etae}), and~(\ref{kappamatrix}).  
For independent parameters we can take $\eta_u+\eta_u^T$, 
$\eta_d+\eta_d^T$, $\epsilon$, $C_1$, and $C_2$.  
Then the symmetric parts of the coefficient matrices 
for the leptons are given by 
\begin{eqnarray}
\eta_e+\eta_e^T &=& -\frac{3r_1+r_4}{r_1-r_4}(\eta_d+\eta_d^T)
               +\frac{4}{r_1-r_4}(\eta_u+\eta_u^T), 
                             \\ 
\eta_\nu+\eta_\nu^T &=& -\frac{4r_1r_4}{r_1-r_4}(\eta_d+\eta_d^T) 
     + \frac{r_1+3r_4}{r_1-r_4}(\eta_u+\eta_u^T), 
                                    \\
M_{\nu_R} &=& \frac{\sqrt{3}v_S}{(C_1)_{41}}
\left[\frac{r_1}{r_1-r_4}(\eta_d+\eta_d^T) 
   -\frac{1}{r_1-r_4}(\eta_u+\eta_u^T)\right],  
                                 \\ 
r_1 &=& \frac{(C_2)_{11}}{(C_1)_{11}}, \quad r_4=\frac{(C_2)_{41}}{(C_1)_{41}}.  
               \nonumber 
\end{eqnarray}
The mass matrix $M_{\nu_R}$ is symmetric.   
The antisymmetric parts of the coefficient matrices are given by 
\begin{eqnarray}
\eta_d-\eta_d^T &=& 2\left [(C_1)_{21}
                  +\frac{1}{\sqrt{3}}(C_1)_{31}\right ]\epsilon,  
          \\ 
\eta_u-\eta_u^T &=& \frac{4}{\sqrt{3}}(C_2)_{31}\epsilon, 
          \\ 
\eta_e-\eta_e^T &=& 2\left [(C_1)_{21}-\sqrt{3}(C_1)_{31}\right ]\epsilon, 
          \\
\eta_\nu-\eta_\nu^T &=& 4(C_2)_{21}\epsilon. 
\end{eqnarray}
Depending on the structures of Eqs. (\ref{higgs_h1}) and (\ref{higgs_h2}), 
the $\bf 120$ representation could contribute exclusively to any one of 
$\eta_d$, $\eta_u$, $\eta_e$, and $\eta_\nu$.  

     We now make an assumption that the 
$(\sqrt{3}H^{\bar 5}_{120}+H^{\overline{45}}_{120})/2$ and $H^{45}_{120}$ 
components in the Higgs superfields $H_1$ and $H_2$, respectively, 
can be neglected, 
taking the equations $(C_1)_{21}+(1/\sqrt{3})(C_1)_{31}=(C_2)_{31}=0$.    
Then, the coefficient matrices for the quarks become symmetric,   
$\eta_d = \eta_d^T$, $\eta_u = \eta_u^T$. 
Adopting a generation basis in which the coefficient matrix for the up-type 
quarks is diagonal, we obtain the equations 
\begin{equation}
\eta_d =V_{CKM}^*\eta^D_dV_{CKM}^\dagger,  \quad  
\eta_u = \eta^D_u.  
\label{quarks}
\end{equation}
The matrices $\eta_d$ and $\eta_u$ are determined by their 
eigenvalues and the CKM matrix.  
On the other hand, 
the coefficient matrices for the leptons are given by 
\begin{eqnarray}
\eta_e &=& -\frac{3r_1+r_4}{r_1-r_4}\eta_d +\frac{4}{r_1-r_4}\eta_u 
            +4\tep,   
\label{electrons}  \\ 
\eta_\nu &=& -\frac{4r_1r_4}{r_1-r_4}\eta_d +\frac{r_1+3r_4}{r_1-r_4}\eta_u 
             + 2r_2\tep,  
\label{neutrinos1}  \\ 
M_{\nu_R} &=& \frac{2\sqrt{3}v_S}{(C_1)_{41}}
          \left[\frac{r_1}{r_1-r_4}\eta_d -\frac{1}{r_1-r_4}\eta_u\right], 
\label{neutrinos2}  \\ 
 \tep &=& (C_1)_{21}\epsilon, \quad  
     r_2=\frac{(C_2)_{21}}{(C_1)_{21}}.  
                   \nonumber   
\end{eqnarray}
The matrices $\eta_e$, $\eta_\nu$, and $M_{\nu_R}$ are expressed as 
linear combinations of $\eta_d$, $\eta_u$, and $\tep$.  
With $\eta_u$ being diagonal, 
the matrix $\eta_d$ is roughly diagonal simultaneously.  
However, for the matrix $\tep$, only off-diagonal elements have 
non-vanishing values.  
The contribution of $\bf 120$ may make the off-diagonal elements 
of $\eta_e$ and/or $\eta_\nu$ non-negligible, which could enhance 
the generation mixing for the leptons.    

     The independent model parameters at the GUT energy scale are given by 
the diagonal matrices $\eta^D_d$ and $\eta^D_u$, the 
CKM matrix $V_{CKM}$, the ratio $r_1$, $r_2$, and $r_4$, 
the antisymmetric matrix $\tep$, and the right-handed 
neutrino mass scale $v_S/(C_1)_{41}$.  
At the electroweak nergy scale, the eigenvalues of 
$\eta_d$, $\eta_u$, and $\eta_e$ are known experimentally, 
if the ratio $\tan\beta$ of the VEVs for $H_1$ and $H_2$ is given.  
The CKM matrix has been measured.     
The quantities obtained experimentally for the neutrinos are  
the mass-squared differences and the mixing angles.  
These observed quantities have to be accommodated by suitable values of 
the model parameters at the GUT energy scale.   

\section{Numerical analyses}

\begin{figure}
\begin{center}
\includegraphics[width=9cm,clip]{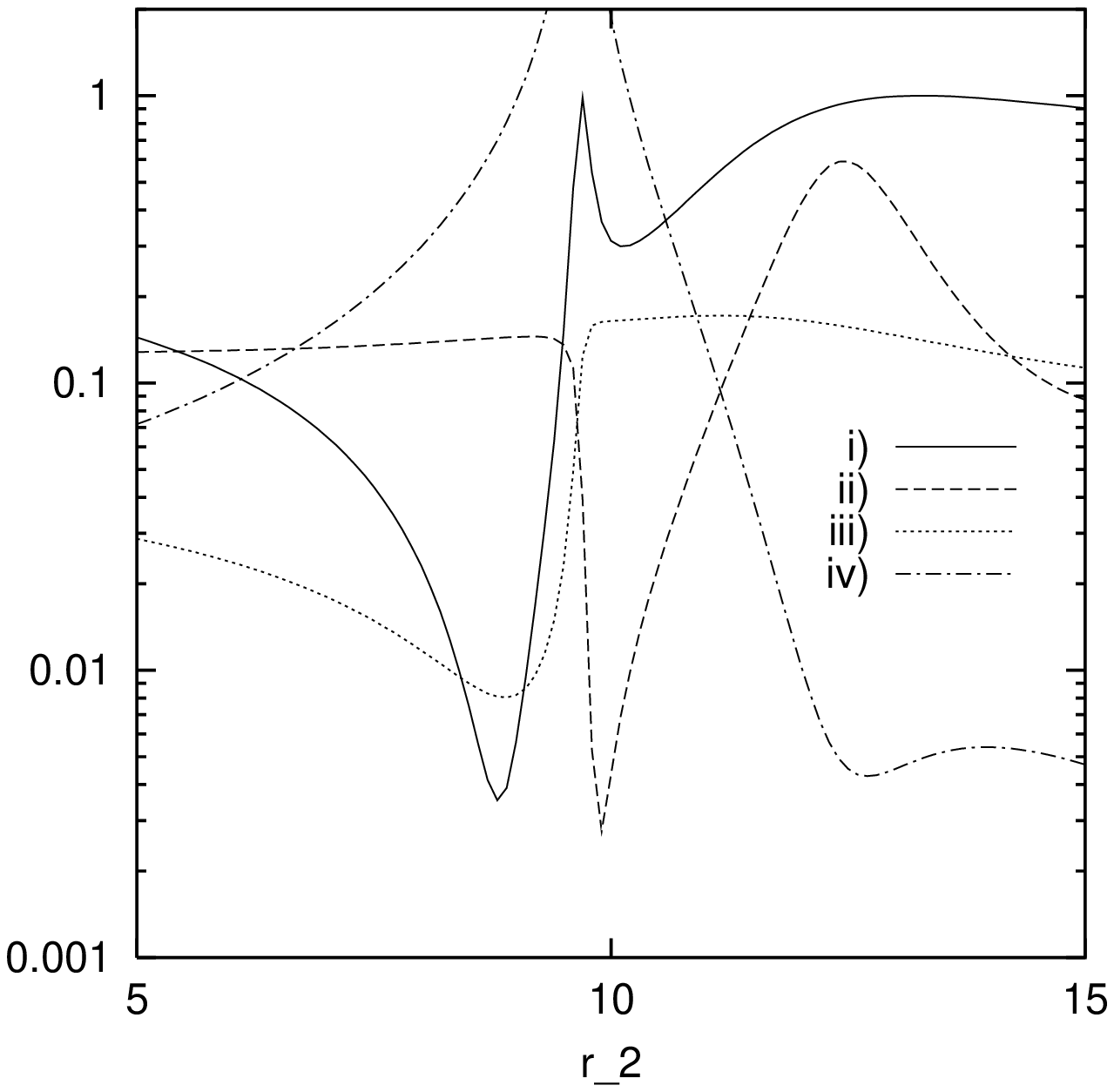}%
\end{center}
 \caption{
The mixing parameters and the ratio of mass-squared differences
for the neutrinos at the electroweak energy scale for the
parameter set (A):
i) $\sinatm$, ii) $\sinsol$, iii) $\sincho$, iv) $\Dsol/\Datm$.
\label{fig:mixA}}
\end{figure}

\begin{figure}
\begin{center}
\includegraphics[width=9cm,clip]{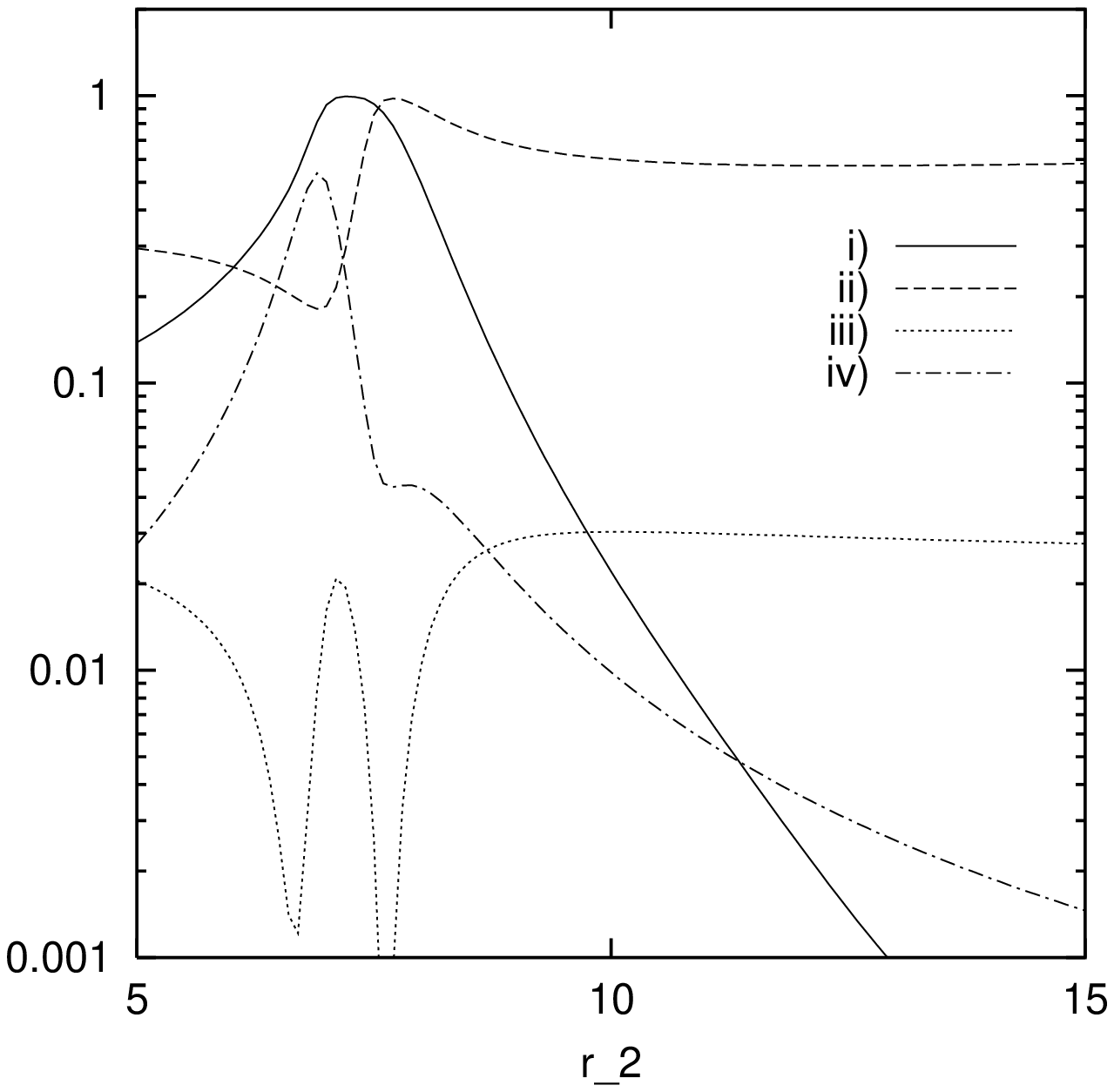}%
\end{center}
 \caption{
The mixing parameters and the ratio of mass-squared differences
for the neutrinos at the electroweak energy scale for the
parameter set (B):
i) $\sinatm$, ii) $\sinsol$, iii) $\sincho$, iv) $\Dsol/\Datm$.
\label{fig:mixB}}
\end{figure}

     We show numerically that the observed neutrino oscillations can be 
described by our model.  Since the number of the model parameters 
at the GUT energy scale is large, a systematical analysis in the 
whole parameter space is complicated.  
Instead, we demonstrate the viability of the model by presenting two 
numerical examples 
within the ranges of real values for $r_1$, $r_2$, $r_4$, and $\tep$.  

     In Figs. \ref{fig:mixA} and \ref{fig:mixB} the mixing 
parameters $\sinatm$, $\sinsol$, $\sincho$, and the ratio of 
mass-squared differences $\Dsol/\Datm$ at the electroweak energy scale 
are shown as functions of the model parameter $r_2$.  
The values of $\tan\beta$, $r_1$, $r_4$, and $\tep$ 
are listed in Table \ref{tab:parameters}, with the sets (A) and (B) 
corresponding to Figs. \ref{fig:mixA} and \ref{fig:mixB}, respectively.   
The other model parameters $\eta^D_u$, $\eta^D_d$, and $V_{CKM}$ 
at the GUT energy scale are tuned to give the quark and charged lepton 
masses and the CKM matrix at the electroweak energy scale shown 
in Table~\ref{tab:massmix}.  
The CKM matrix can be expressed by the standard parametrization \cite{PDG} 
and the 
$CP$-violating phase of this parametrization is denoted by $\delta_{SP}$.   
These masses at the electroweak energy scale for the quarks and charged 
leptons, as well as the magnitudes of the CKM matrix elements, 
are consistent with the experiments~\cite{fusaoka}.  
The $CP$-violating phase $\delta_{SP}$ lies in the range allowed 
by all the $CP$ violation phenomena observed in the $K^0$-$\bar K^0$ and 
$B^0$-$\bar B^0$ systems~\cite{oshimo_bb}.  
The outcomes in Table \ref{tab:massmix} do not vary with $r_2$.  
The value of $v_S/(C_1)_{41}$ determines the scale of $\kappa$ and does 
not affect the presented four observables for the neutrino oscillations.  

\begin{table}
\caption{
The parameter sets (A) and (B) for
Figs. \ref{fig:mixA} and \ref{fig:mixB}, respectively.
\label{tab:parameters}
}
\begin{center}
\begin{tabular}{ccc}
\hline
             &  (A)   &   (B)        \\
\hline
$\tan\beta$    &   20    &    30      \\
$r_1$          &  $-3.9$   &   $-1.9$     \\
$r_4$          &  $-7.3$   &   $-5.0$      \\
$\tep_{12}$ & $-4.0\times 10^{-4}$ & $-9.0\times 10^{-4}$   \\
$\tep_{13}$ & 3.6$\times 10^{-3}$ & 4.9$\times 10^{-3}$   \\
$\tep_{23}$ & 9.2$\times 10^{-3}$ & 7.6$\times 10^{-3}$   \\
\hline
\end{tabular}
\end{center}
\end{table}

\begin{table}
\caption{
The masses of the quarks and charged leptons (in unit of GeV) and
the CKM matrix (its elements and $CP$-violating phase in the
standard parametrization) at the electroweak energy scale.
\label{tab:massmix}
}
\begin{center}
\begin{tabular}{ccccc}
\hline
$m_t$    & 1.8$\times 10^{2}$  & & $|(V_{CKM})_{12}|$ & 0.22 \\
$m_c$    & 6.8$\times 10^{-1}$ & & $|(V_{CKM})_{13}|$ & 0.0036 \\
$m_u$    & 2$\times 10^{-3}$   & & $|(V_{CKM})_{23}|$ & 0.041 \\
$m_b$    & 3.0                 & & $\delta_{SP}$      & 1.3 \\
$m_s$    & 9.3$\times 10^{-2}$ & &   &   \\
$m_d$    & 5$\times 10^{-3}$   & &   &   \\
$m_\tau$ & 1.8                 & &   &   \\
$m_\mu$  & 1.0$\times 10^{-1}$ & &   &   \\
$m_e$    & 5$\times 10^{-4}$   & &   &   \\
\hline
\end{tabular}
\end{center}
\end{table}

     We can see from Figs. \ref{fig:mixA} and \ref{fig:mixB} that 
the atmospheric and solar neutrino oscillations, under the constraints 
from the CHOOZ experiment, are realized simultaneously for certain 
parameter values.   
In Table \ref{tab:mixings} the resultant values of 
$\sinatm$, $\sinsol$, $\sincho$, and $\Dsol/\Datm$ are explicitly 
given for the set (A) with $r_2=12.5$ and for the set (B) with $r_2=7.5$.    
If the mass scales of the right-handed neutrinos are put at 
$v_S/(C_1)_{41}=1.0\times 10^{15}$ GeV in the set (A) and 
at $v_S/(C_1)_{41}=6.0\times 10^{14}$ GeV in the set (B),   
the mass-squared differences are given by   
$\Dsol=2.5\times 10^{-5}$ GeV$^2$, $\Datm=5.5\times 10^{-3}$~GeV$^2$ 
and by  
$\Dsol=8.7\times 10^{-5}$ GeV$^2$, $\Datm=1.6\times 10^{-3}$~GeV$^2$, 
respectively.  
Note that the mixing angle $\sincho$ is predicted to be around the present 
experimental bound for the case (A), while to be much smaller than it for 
the case (B).  

\begin{table}
\caption{
The mixing parameters and the ratio of mass-squared differences
for the neutrinos at the electroweak energy scale with $r_2=12.5$
and $r_2=7.5$ for the sets (A) and~(B), respectively.
\label{tab:mixings}
}
\begin{center}
\begin{tabular}{ccc}
\hline
             &  (A)   &   (B)     \\
\hline
 $\sinatm$   &  0.95  &  0.94   \\
 $\sinsol$   &   0.59  &  0.86   \\
 $\sincho$   &  0.16  &  0.0026  \\
 $\Dsol/\Datm$  & 0.0045  &  0.055  \\
\hline
\end{tabular}
\end{center}
\end{table}

\section{Conclusions}

     We have discussed the masses and mixings of quarks and leptons within 
the framework of GUT coupled to supersymmetry.  
In GUT models the Higgs couplings for quarks and leptons are 
closely related to each other.  
The large generation mixing for the leptons, which is observed 
experimentally through the neutrino oscillations, 
cannot coexist trivially with the small generation mixing for the quarks.  
Some natural explanation for the difference of mixing between the quarks 
and the leptons is sought.   

     To solve the problem of generation mixing, 
we have proposed a model with SO(10) and supersymmetry.  
This model is a simple extension of the minimal SO(10) model, 
and includes a Higgs superfield of $\bf 120$ representation, 
as well as two Higgs superfields of $\bf 10$ and $\overline {\bf 126}$
representations.  
The energy evolution of the Higgs couplings with quarks and leptons 
have also been taken into account between the energy scales 
of GUT and electroweak theory.  

     By the simple enlargement of the Higgs sector, 
the masses and mixings of the quarks and leptons are 
consistently accommodated without invoking much contrived schemes.  
The $\bf 120$ representation makes the Higgs couplings 
different between the quarks and the leptons.   
The small neutrino masses are traced back to 
large masses for the right-handed neutrinos and sneutrinos 
generated by $\overline {\bf 126}$.  
The model parameters are constrained to correctly give the quark 
and charged lepton masses and the CKM matrix.  
In spite of these constraints, the experimental results for the 
neutrino masses and the MNS matrix can be described well  
in certain regions of the parameter space.  

\ack{}
     This work is supported in part by the Grant-in-Aid for
Scientific Research on Priority Areas (No. 14039204) from the
Ministry of Education, Science and Culture, Japan.


\end{document}